# The Object Oriented Approach to Control Applications and Machine Physics Calculations with Java Technology

M. Kadunc, I. Kriznar, M. Plesko, G.Tkacik, J. Stefan Institute, Ljubljana, Slovenia

Abstract

At least as much time is spent solving problems addressed by application support as with actual interaction with the control system, if it is not factored into separate libraries but is coded by each programmer time and time again. Our Java libraries, called Abeans, address these issues in a communication platform independent way. Abeans perform connection management, error and timeout handling, logging, reporting, authentication, resource initialization and destruction, policy and configuration management, etc. Abeans also hide the communication details from the programmer and combine with CosyBeans, our GUI Java components, to form SCADA-like control panels. They are successfully used in such different environments such as a synchrotron light source (ANKA, Germany), an e-p collider complex (DESY, Germany), proton cyclotron (Riken, Japan) and a radio telescope (ESO, Germany).

The Java library DataBush has a similar function for machine physics programs. The object oriented design of the library gives programmer intuitive access to devices and elements relevant to machine physics. For example, a non-visual Java Bean represents a magnet with its relevant machine physics parameters. Access to the control system is provided transparently with Abeans in communication platform independent way. Error diagnostics and event handling is part of the Abeans and DataBush framework. By default DataBush receives new data and performs linear optics calculation of machine functions with a one second heartbeat.

## 1 INTRODUCTION

The efforts of the authors of this paper in the last decade have been always directed towards creating a simple and intuitive access to control system functions. From the experiences with the databush C library developed at Elettra [1] for machine physics applications, it became clear that a general approach that hides the underlying communication from the client and presents controlled devices as objects is a viable way to go. For clarity, we have separated the tasks into two libraries.

**Databush** [2] transforms data from Abeans into machine physics parameters (e.g. power supply current to focal strength of quadrupole) and vice versa and provides machine physics calculations.

The **Abeans** [3] library is used by any application, not just for machine physics. It

- Provides transparent access to controlled devices by wrapping the underlying communication mechanism through plugs
- Allows GUI widgets to connect to the control system in rapid application development tools
- Provides application support through a series of services that should cover nearly all application types and their problems.

Abeans have a pluggable model of control systems, which means they are applicable to any control system from EPICS, CDEV, TINE, TANGO, and ACS to proprietary ones. Only the appropriate plug has to be written, but the application does not change. Both libraries have a series of support functions that discover devices at run-time, sort and rearrange the objects by type, handle errors, etc.

The DataBush reads values from the control system transparently via Abeans. DataBush also makes it possible to send values to the control system directly as machine physics properties. The accelerator specific implementation is hidden in machine physics configuration files and the control system specific implementation in the Abeans. Two major benefits come out of this architecture:

- **Portability**: Control and machine physics applications can be used on other accelerators without changing a single line of code, with appropriate modification of the Abeans plugs of course. This enables sharing accelerator software among accelerator centers.
- **More efficient programming**: An operator or physicist can concentrate on machine problems; no special knowledge of control system is needed, since DataBush and Abeans transparently handle all communication with the control system.

## 2 DECIDING FOR JAVA BEANS

It was only natural to switch from C to an object oriented language, which provides the proper basis for objects. We have chosen Java because it is a modern object oriented programming language, it has well defined data types and API (Application Programming Interface), it allows easy use of graphic widgets, threads and other system tools without having to know the specifics of a given platform. Java is also an interpreted language, so it is a little slower than compiled languages like C++, but we found out that by using JIT (Just-In-Time) compilers it is fast enough for our needs. The fact that Java is platform independent makes our approach potentially available to any control system.

A Java bean is a reusable component written in Java that can be manipulated in a visual builder environment, similar to GUI-builders or Visual Basic: beans can be graphically arranged and connections between them established. Such environments, commercially available at a reasonable price, enable the programmer to build an application without typing a single line of code. Many GUI beans exist, such as labels, buttons, gauges, charts, etc. However, beans can be also "invisible", having pure functionality without graphical representation.

Such invisible beans are contained in the Abeans and Databush libraries. For each device type there is one corresponding device bean, which behaves to the programmer as if it were the device itself. A device bean encapsulates all remote calls from the client to a device server, e.g. get/set, on/off etc. Thus the network is invisible to the user of device beans. Once the Abeans wrap the control system, they can perform further tasks that an application programmer has to provide for each application: Tasks of a device bean include opening the connection and performing the function calls on remote objects; report and manage all errors/exceptions/timeouts, providing handles for asynchronous messages and the like. Thus a single asynchronous update of a power supply current travels through the whole structure and results in an automatic update of the machine physics parameters such as beta functions and tune.

The design patterns of Abeans and Databush are sufficiently mature so that they can be implemented also in another programming language. However, the advantage of using visual builders as for Java Beans is lost in that case.

## 3 DATABUSH

The following features make DataBush a flexible machine physics tool
- Quasi real-time current or magnetic properties transformation driven by events from the control system
- Quasi real-time linear optics calculation of machine functions
- Automation of often-used machine physics routines
- Representation of the machine with Java beans
- Type- and cast-safe programming with DataBush beans
- Customization of DataBush operations with pluggable mechanisms

The main DataBush design guideline is to
- **Make it flexible and adjustable to particular needs of CS and machine physics application implementation**. For this DataBush itself has no visual presentation, however several visual beans were developed displaying most important properties (tune, machine functions, and orbit...). There are also some pluggable interfaces, with which default implementation can be replaced with more suitable to particular implementation.
- **Make a run-time data structure that presents accelerator lattice in consistent object-oriented way**, to use advantages of Java language. This means that each lattice element has it's own Java bean - a DataBush element. Every element has certain properties and methods concerning its physical contents.

DataBush elements are presenting the following three main types: magnets (influence on accelerated particles), beam position monitors (diagnostic of particles) and power supplies (control magnets) and some other special elements (table 1). Each element has properties and methods consistent with its nature. Elements that are related, for example all magnets, have similar behavior and properties. To reuse code, related elements are extended from the same ancestor bean, which encapsulates their common behavior. This can be seen in figure 1.

Table 1: Short summary of DataBush elements with description

| DataBushInfo | holds energy and some other key properties concerning all elements in DataBush |
|---|---|
| PowerSupply | holds current, supplied to magnets |
| BPMonitor | BPM, holds position of electron beam |
| Marker | Special element kind, that just marks selected position on optical path |
| Aperture | Collimator with its hole dimension |
| Cavity | Cavity, radio frequency element |
| RBending | Rectangular dipole magnet, holds dipole field, radius |
| SBending | Sector dipole magnet, holds dipole field, radius |
| HorCorrector | horizontal corrector (steerer), holds kick |
| VerCorrector | vertical corrector (steerer), holds kick |
| Kicker | Steerer for kicking the electrons around injection point |
| Quadrupole | quadrupole magnet, holds quadrupole field momentum |
| Sextupole | Sextupole magnet, holds sextupole magnet momentum |
| Septum | Septum magnet |

Elements are organized in lists that are based on Java Collections. Each main group of similar elements has its own list with its own list iterator. This makes work with DataBush type- and cast-safe. Organization of DataBush devices with list, and additional references between elements allows browsing the elements forward and backward, searching neighboring elements in all sorts of relationships. Such unique devices relationships make a bush like structure and from this comes the package name DataBush.

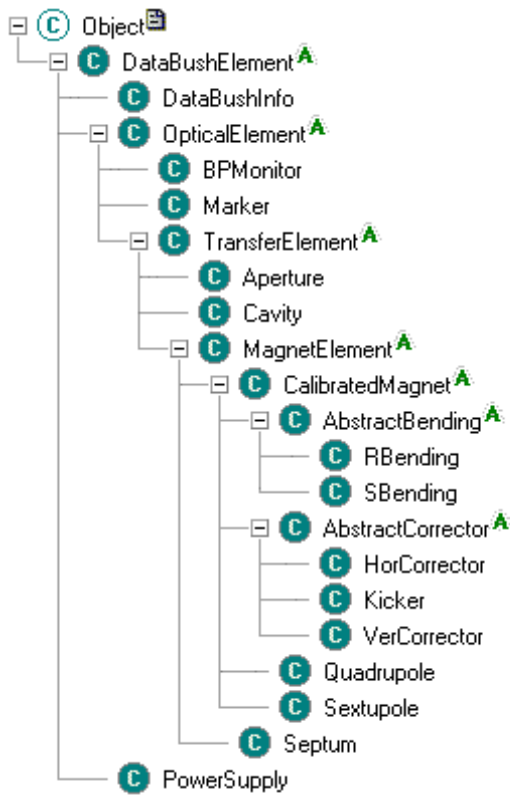

Figure 1: Diagram of element inheritance. All DataBush elements have one common ancestor, the DataBushElement.

Any DataBush element can be marked as simulated. Such an element does not connect to the underlying control system and just holds values in read/write manner. With this feature only part or whole DataBush can be used as simulation of the real machine. With simulated DataBush, a machine physics application can be tested before it is used on the machine. The DataBush simulation mode can also be used as an ordinary machine physics calculation program for linear optics calculations.

## 4 ABEANS

Abeans are Java libraries consisting of graphical and process logic components that assist programmers in developing client applications for control systems. They evolved from simple wrappers used for hiding complex details of communication protocols like CORBA into full-featured set of tools providing the following functionality:

**Java beans models of control system**: Each physical device is represented by a device bean. Each property of a device is represented by a property bean. Java beans are special Java classes that conform to Sun Java Beans specification. Such classes are 100% pure Java and can be used in any integrated development environment for Rapid Application Development (RAD), such as IBM Visual Age for Java, Inprise JBuilder, Symantec Visual Café etc.

**Application building libraries:** These include common services needed by application developers, such as logging, reporting, error and exception handling, configuration management, lifecycle management and the like.

**A set of GUI widgets** called CosyBeans tailored specially for displaying physical values and their context. For example, an enhanced version of Swing slider is available with additional features, such as display of units, display format specification, minimal step setting, bit-by-bit increment / decrement, different setting modes (synchronous, non-blocking), periodic value update from remote control system data source etc.

By combining the features mentioned above, the application developer can approach two areas of interest listed below with ease:

**Simple GUI panels**: These panels consist of GUI widgets that connect to specific devices in control system in order to display and set the controlled values. The development of simple panels is possible in visual composition editors by simply dragging device beans from the palette onto the free form surface. The following step consists of drawing the GUI by positioning and customizing GUI widgets. Finally, device beans and GUI widgets are wired together by connecting them visually with lines that denote data flow from the device into the widget and vice versa. After such visual composition is complete, the integrated development environment will automatically generate Java source code for the application. Abeans thus truly deliver what RAD paradigm promises: functional applications without a single line of code written by hand.

**Complex control applications**: To fulfill the need for advanced applications that query the data from many devices and send a lot of requests, Abeans libraries can also be used as ordinary Java classes in manual programming. In this case, the object-oriented design of Abeans provides an unmatched clarity of control system concepts, reducing the time needed for programmers to become truly productive. The Abeans libraries automatically determine that they are being used manually (as opposed to being used in visual composition builders) and change the default behavior to match the programmers' intuitive expectations of the library semantics.

An additional powerful feature of Abeans is the pluggable architecture. While exposing a uniform, consistent, object-oriented and intuitive interface to the application programmer, the Abeans can plug into many different communication mechanisms used by remote servers. Abeans themselves provide a Simulator plug, which creates "virtual" devices on the local machine, allowing users to code Abeans applications at home without the need to actually access the real remote processes. When the application is complete, the developer easily switches (even at run-time) from Simulator plug to a real (let's say CORBA or TINE protocol) plug and the application talks to real servers instead of virtual ones.

This technology has enabled us to quickly port Abeans from initial implementation with Visigenic CORBA plug in ANKA to Orbacus CORBA plug for ESO and to make plugs for completely different kinds of communication models such as TINE at DESY [4] or prototype models at RIKEN [5]. In principle it is even possible to run the applications developed for one machine on a different one, just by switching the plugs!

Currently under development is a complete rewrite of Abeans libraries, called Abeans R3 (short for Release 3). Whereas existing Abeans R2 provide a stable and tested platform for application development, the new release will incorporate all experiences gathered during the ANKA project as well as comments and requests from other institutions that tested Abeans. Abeans R3 will provide additional support for application building, including debugging support and extensive exception handling. Control system models will be pluggable in the next release: the developer will be able to view the control system either as a collection of devices containing properties (wide interface), as a remote database, or as a collection of data channels (narrow interface) - or for that matter any other suitable model. Needless to say, all of the benefits of Abeans R2 will be integrated into the new release.

## 5 COSYBEANS

To represent the values of the control variables (pointing vector, status of the Mount, etc.), several GUI components were developed under the general name CosyBeans: gauger, slider, trend, ledder, textPane, WindowFrame, inputField, wizard, etc. They were specifically designed to represent the properties of controlled devices and are therefore tailored to be used with Abeans. It is even possible to create simple applications in a visual builder environment without typing a single line of code: all that needs to be done is to connect a property (e.g. current) from an Abean that represents a certain device (e.g. a power supply) to a corresponding GUI component (e.g. gauger). Another component is the selector, which enables the user to search for all available devices of a given type dynamically at run time and chose one or a group of them. When the choice is made, the Abeans automatically take care of the initialization process and the gauger is immediately showing the correct value. An example of such an application is shown in figure 2.

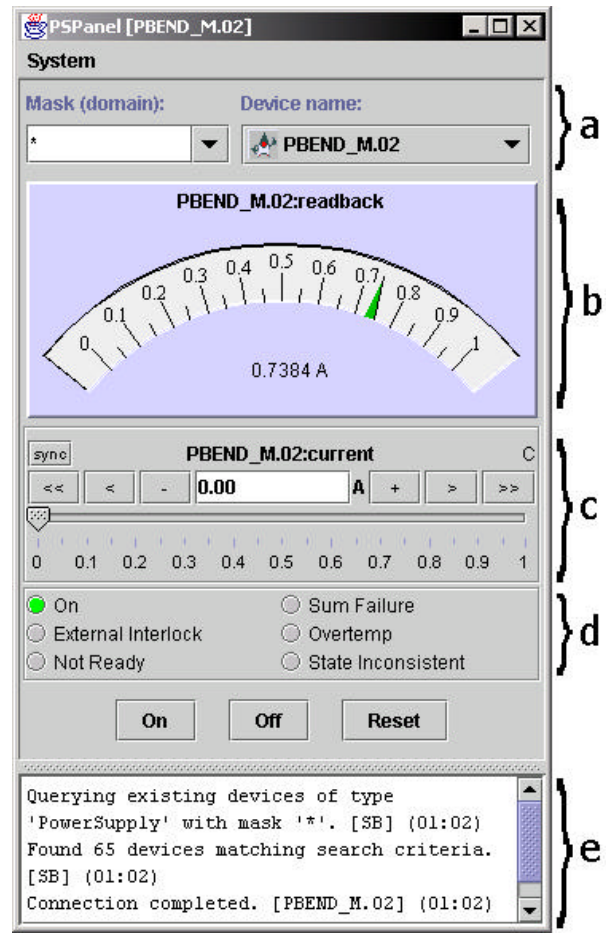

Figure2: Power Supply Panel. This panel was created from Abeans and GUI components in Java visual builder environment without typing a single line of code. The following GUI components can be seen: a) selector, b) gauger, c) slider, d) ledder, e) ATextPane - a pane with messages for the user.

Great care has been taken in the design of the GUI components. They don't just display a single value, but provide visualization for all characteristics of a property. Each gauge or slider can spawn a trend chart. Limits and display precision is read from the property by default but can be set by the user at run-time. Among others, keyboard shortcuts, logarithmic scale, different modes of value get/set, different refresh rates, tool-tip text and alarms are all supported. The trend chart has variable history length; supports zoom and pan, saves data in TAB-delimited format and can convert to a histogram chart.

## 5 DETAILS OF ABEANS R3

Historically, Abeans were developed as a client side of the control system for ANKA synchrotron light source. They provided wrappers for CORBA that communicated with remote servers and hid the complexity of CORBA from the average Java GUI programmer. Since then, Abeans evolved several distinguishing features, like pluggable communication protocol, enabling Abeans to talk to non-CORBA servers, a set of specialized GUI widgets appropriate for display of physical variables and the like. At the same time, Abeans provide several useful services: logging, exception handling, configuration and data resource loaders. However, the design legacy that limited Abeans to the domain of wide-interface control systems remained. Designs of version 3 of the Abeans release, called R3, are a result of the need to address the following issues (check [6] for newest documentation):

· Provide libraries for application development irrespective of whether they interact with control system or any other data source. Minimize the need for ANY repetitive code in client applications.

· In addition to pluggable communication protocols provide also pluggable models of control system. For example, enable the programmers to view the control system as a collection of Devices containing Properties or as a collection of Channels.

· Provide both narrow- and wide-interface access to the control system. Provide both programmatic (as in procedural programming) and declarative (as in XML descriptions) ways to access the control system.

· Provide pluggable architecture where the implementations of different services can be switched, preferably at run-time (e.g. change logging implementation while the client is running).

· Enforce strict separation (conceptual and programmatic) between models, plugs, core libraries and libraries for visualization (GUI widgets).

In Abeans R3, different models are used to represent the structure of the control system. Models use plugs to get data from a specific control system. At DESY for example, we used the Abeans "channel" model (i.e. a narrow interface access model), which consists of namespaces and channels, to create a plug connecting the TINE Java class to the Abeans.

As Abeans R3 are designed to run multiple applications in a single JVM (Java Virtual Machine), libraries are loaded only once and the memory footprint is modest compared to applications running in separate JVMs. This feature also allows the same applications to be run individually or from within an applet in a web browser.

These are only two of over 30 new features that are already implemented in R3.

The structure of Abeans is modular: the classes separate into four groups of packages: Core libraries, framework libraries, core modeling library and core plug library are fixed and form a static portion of Abeans. Other packages are pluggable, i.e. contain implementations of the interfaces declared in the fixed portion of Abeans. A minimal working set of Abeans consists of the fixed part plus at least one implementation for each required pluggable component. Implementations of specific models are delivered as modeling libraries that extend the core modeling library. Communication protocols between remote processes and a given model are delivered as plug libraries. Core libraries prescribe interfaces, abstract base classes and design patterns that all pluggable libraries must conform to.

## 4 CONCLUSIONS

Abeans run now at several places. This justifies further development and we plan to finish release 3 within a few months, which should provide solutions to the problems of application support once and for all. The question remains, whether a Java library is acceptable for all developers, or would there be need for a C++ version?

## 4 ACKNOWLEDGMENTS

We thank the institutes DESY, ESO, FZK, Riken and the SLS for hosting us and showing continuous support for our ideas.